# Influence of Personality Types in Software Tasks Choices

Luiz Fernando Capretz, Western University, London, Ontario, Canada - N6G5B9, lcapretz@uwo.ca
Daniel Varona, University of Informatics Science, Havana, Cuba, dvarona@uci.cu
Arif Raza, Western University, London, Ontario, Canada - N6G5B9, araza22@uwo.ca

*Abstract*— According to psychology, not everybody can excel at all kinds of tasks. Thus, chances of a successful outcome of software development increase if people with particular personality traits are assigned to their preferred tasks in the project. Likewise, software development depends significantly on how software practitioners perform their tasks. This empirical study surveys 100 Cuban software developers, both students and professors of the University of Informatics Sciences in Havana, Cuba. This work aims to find possible patterns that link personality traits to role preferences in a software life cycle. Among the various roles, system analyst, software designer, and programmer are found to be the most preferred among the participants. In contrast, tester and maintainer happen to be the least popular roles among software engineers.

## I. INTRODUCTION AND BACKGROUND

Software engineering has been one of the most prominent professions over the last 20 years, and it is projected to evolve even further. Software engineering comprises stages in distinct areas, such as analysis, design, programming, testing, and maintenance. Today, specialties within software engineering are as diverse as in any other profession. Additionally, software engineers need to communicate more effectively with users and team members, thus the people dimension of software engineering is as important as technical expertise.

Software project managers have always faced the problem of assigning the tasks to the right people within a team in such a fashion that increases the chance of successful project completion [1]. Different ideas have been tried to use diverse ways to maximize performance [2] and make choices in the software engineering process [3]. Those ideas involve: motivation (software engineers tend to perform better if they are motivated to do specific tasks), the environment, and personality type, or a combination of these factors. Motivation and the environment are known to influence task performance. Motivation is generally a powerful element in the performance of task goals, especially in the IT field [4], [5]; however, motivation is often insufficient for influencing performance of goals on its own. Similarly, environmental factors cannot independently generate the performance of tasks. Hence, there are multiple factors involved in the performances of software engineers [6]. This study specifically investigates the role of individual preferences in software projects, while neglecting the elements of motivation and environment, which have been the focus of most scholarly research on this topic. Feldt et al. [7] also state that environmental factors alone cannot improve task performance. Thus this work exclusively investigates the role of individual preferences in software projects, focusing explicitly on how personality types affect preferences for specific software roles.

Several studies investigate the relationship between software engineer personalities and performance by identifying associations between particular personality types and specific tasks in software development. For example Choi [8] and Da Cunha [9] report specific issues related to programming. Acuna and Juristo [10] introduce a capability-person relationship model that can be used by software project managers to assign tasks to people based on soft skills. Acuna et al. [11] report that properly assigning people to development roles is crucial for creating productive teams, and their human capacity-based procedure can aid managers at small- to medium-sized software organizations.

Ritcher and Dumke [12] adapt the Big Five method for software engineering with a Failure Mode and Effect Analysis method that models the human factor as a risk factor in the software engineering process and examines methods to evaluate psychological characteristics to diagnose expected productivity. Capretz and Ahmed [13] present a better understanding of the general preferences of software engineers in software life cycle phases and map these phases to the Myers-Briggs Type Indicator (MBTI) dimensions taking into consideration desirable soft skills that appear in job ads. As far as these studies are concerned, not only performance and task choices are affected by personality type, but also other factors, such as motivation and the surrounding environment, can influence task choice, motivation to carry out a task, and task performance.

A wide variety of psychological instruments are used for career counseling and behavior prediction. In understanding the influence of personality on software development tasks there exists a wide variety of personality frameworks (e.g., Five-Factor theory, Keirsey Temperament Sorter, etc.). The MBTI [14] is one of the most popular tools used in workplaces to analyze personality types. According to the MBTI, a person is measured across four dimensions by his/her preferences: energizing, attending, deciding, and living.

Within each dimension, there are two opposite poles: Extroversion (E) – Introversion (I), Sensing (S) – Intuition (N), Feeling (F) – Thinking (T), and Perceiving (P) – Judging (J). Sixteen distinct personality types are defined on the basis of combining these preferences; each type is denoted by four letters. These distinctions have an influence on career choice because people tend to choose occupations that are related to their personality type.

Within the first dimension, Extroverts get their energy from interactions with people, are outgoing, and prefer to work with other people, whereas Introverts get their energy internally and prefer to work alone.  Secondly, the S-N dimension is related to the way in which people acquire information. In particular, sensing people receive information from their five senses and are attuned to the practical, hands-on, common-sense approach to information; intuitive individuals are more focused on complex interactions, theoretical implications, and new possibilities. The third dimension, T-F, is concerned with how people make decisions. Specifically, thinkers prefer to analyze logical/objective data. In contrast, feelers respond to situations depending on their feelings about that situation and often want work that provides services to people. Finally, the Judging type prefers work that has a need for order, whereas the Perceiving type prefers tasks that require adapting to changing situations.

The MBTI has its critics [15] who point out shortcomings with its statistical structure and other limitations [16]. We should be cautious about its possible misuse in organizational and occupational settings. However, MBTI continues to be the most popular instrument used in profiling the personality types of software engineers [17].

Myers [14] shows that an individual's interest in jobs is mainly determined by the S-N and T-F dimensions. These pairs are responsible for the cognitive scales that influence the extent to which people feel attracted to and satisfied by their career choices: STs prefer activities that require the use of established knowledge and are observant and detail-orientated, they are reluctant to try new innovative solutions; NTs are creative and, consequently, enjoy symbolic abstract relations and seek to find patterns rather than dealing with details. Additionally, they like to create new knowledge rather than applying or improving existing techniques. NTs are more creative than STs because Ns see possibilities beyond the given facts and look for patterns and relationships. Thus, when NTs join both theoretical frameworks with their tendency to extrapolate beyond the details, they can identify new principles. The extroversion-introversion and judgment-perception dimensions determine individuals' personal attitudes.

Most studies concerning the MBTI distribution among students and engineering professionals demonstrate that ISTJ, INTP, and ESTJ are over-represented personality types, whereas ENFJ and INFJ types are underrepresented [18]. The personality distribution of computer application software engineers can be seen in Table I, which presents data taken from the book *MBTI Type Tables for Occupations* [19] for Software Engineers, System Analysts, and Programmers.

Cruz et al. [20] present a comprehensive systematic literature review of personality in software engineering. Other researchers have studied characteristics and traits in personality types for certain roles in software engineering [21], [22], [23], and [24]. System Analysts and Programmers are among the more explored roles in these studies. Nevertheless, we do not find a straight relationship among their preferences and personality types. Since there is a logical relationship between the task preferences and proportions of personality types in software engineering, our study seeks to provide a precise configuration of relationships between personality types, task preferences, and roles in order to obtain conclusive results. In particular, an empirically validated study of actual software developers is used to investigate these relationships.

## II. MOTIVATION

As a discipline, software engineering consists of many roles and responsibilities from the perspective of a project team.

The definition of roles significantly depends on the project characteristics and the development process. While there are a wide range of roles in software development, this investigation focuses on some defined roles: analyst, designer, programmer, tester, and maintainer [25].

The MBTI type distribution among software professionals is presented in Table I [19].

TABLE I - MBTI Type Distribution among Software Engineers, System Analysts and Programmers [19]

| ISTJ | ISFJ | INFJ | INTJ | E | I |
|---|---|---|---|---|---|
| se 17.3% | se 3.6% | se 2.2% | se 9.0% | se 42.8% | se 57.2% |
| sa 17.7% | sa 4.8% | sa 2.0% | sa 6.7% | sa 48.7% | sa 51.3% |
| p 19.4% | p 5.0% | p 2.6% | p 7.6% | p 38.5% | p 61.5% |
| ISTP | ISFP | INFP | INTP | S | N |
| se 8.1% | se 1.6% | se 3.9% | se 11.5% | se 52.0% | se 48.0% |
| sa 5.7% | sa 3.0% | sa 4.3% | sa 7.1% | sa 57.9% | sa 42.1% |
| p 9.1% | p 3.3% | p 5.4% | p 9.1% | p 58.3% | p 41.7% |
| ESTP | ESFP | ENFP | ENTP | T | F |
| se 4.7% | se 2.0% | se 3.8% | se 9.7% | se 78.9% | se 21.1% |
| sa 5.6% | sa 2.3% | sa 4.8% | sa 7.1% | sa 71.9% | sa 28.1% |
| p 5.0% | p 2.1% | p 4.4% | p 5.4% | p 71.4% | p 28.6% |
| ESTJ | ESFJ | ENFJ | ENTJ | J | P |
| se 12.7% | se 2.1% | se 2.0% | se 6.0% | se 54.8% | se 45.2% |
| sa 14.1% | sa 4.7% | sa 2.2% | sa 7.9% | sa 60.1% | sa 39.9% |
| p 9.9% | p 4.5% | p 1.3% | p 5.9% | p 56.2% | p 43.8% |

Note 1: se means "software engineers," sa means "system analysts," and p mean "programmers".
Note 2: Sample of: 1,326 subjects for se, 2,493 subjects for sa, 1,719 subjects for p.

Software engineers review, design, create, and test software for basic computer applications, including operating systems, compilers, and computer networks. They expand existing or launch new general software applications and may also examine or design databases. They establish operational specifications and study requirements using computer science, engineering, or mathematics. The software engineering discipline can be seen as an umbrella career for specialized tasks.

System analysts must be able to understand system essentials, and to create an abstract model of the application in which user needs are met. They also require a significant amount of interaction with users/clients to understand the scope and constraints of the software solution, and to propose procedures to improve software capabilities and workflow.

Software designers are responsible for modeling the proposed system as well as evaluating and validating the designed model to ensure its quality before generating the software code. According to Shatnawi and Alzu'bi [26], unlike other engineering products software systems are not tangible. Due to their complexity, software quality assurance is essential to maintain its quality. Nevertheless, designers identify appropriate components by experimenting with a variety of schemes in order to discover the optimal way of refining the application for the client. Ideally, software designers should have sufficient skills for producing a model that supports the programmer activities and provides an accurate translation of the customer requirements into a finished software product that is stable and operates on real machines.

Programmers transfer project specifications into detailed algorithms for coding into a computer language. They create and write computer programs to perform calculations and store, locate, or retrieve data or other information. The process of programming requires the identification of control structures, relevant variables, and data structures as well as a detailed understanding of the syntax and specifcs of a programming language. In particular, programmers need to attend to details and maintain an open, logical, analytical style. MBTI type distribution among computer programmers is also presented in Table I [19].

The task of testing is focused on identifying faults; there are many ways to make testing efforts more efficient and effective. Testing strategies are neither random nor haphazard; rather, they should be approached in a methodical and systematic manner. The process of debugging errors can be a frustrating and emotionally challenging activity that can lead software engineers to restructure their thinking and decisions. Takamatsu et al. [27] realize that software testing is costly. They recommend automated testing and consider it to be an effective way to reduce the burden of testing. However, testing requires persistence, especially because of the need to choose from an enormous range of possibilities and maintain a high level of attention to detail.

Software maintenance involves keeping applications operational, reacting quickly to problems when produced in order to restore service, meeting or exceeding the agreed level of service, and maintaining the confidence of the user community. Specifically, maintenance personnel should ensure that users believe in their support team as dedicated and competent individuals who are acting within the agreed budget. Unfortunately, we could not find data providing MBTI type distribution among designers, testers, or maintainers to do comparisons with observed results.

## III. RESEARCH METHODOLOGY

This paper aims to identify software engineers' preferred roles on the basis of their distinct personality types. We have used an empirical method to relate personality types and preferences in software engineering roles. This constitutes an important link between personality types and tasks to which engineers are inclined. In order to understand whether individual personality affects software engineers' performance, specific results from the field are required. Consequently, 100 software developers from the University of Informatics Sciences (UCI) in Havana, Cuba, have been surveyed. For the sake of clarification, UCI is an atypical university, where both students and professors are directly engaged in software projects. Their average experience as software developers is five years. So it is correct to think of UCI students/professors as software practitioners, as they develop software products that are exported to several countries.

In this study, 100 Cuban software developers were surveyed, including students in senior Informatics Sciences Engineering courses and professors at the University of Informatics Sciences in Havana, Cuba. There were seven senior students and 93 professors among the subjects. Both students and professors were directly involved in software development with an average experience of five years as software developers. So it is accurate to think of UCI students and professors as software practitioners. Actually, they develop software products that are exported to several Latin American countries and a few European countries. They were invited to take part in the survey based on their willingness to do so. The sample contained 47% males and 53% females. It is the country's policy that all Cuban universities have a balanced ratio between males and females to avoid gender bias. The student age range was between 22 and 23 years of age, while professors ranged from 23 to 27 years of age. The MBTI instrument, Form M, Spanish language version, was used to identify their personality types. Although this instrument is self-assessed, a CPP Certified evaluator was present and processed the data.

As far as the individual roles were concerned, prior to collecting this information about role preferences, the subjects were given the role definitions. The seven students in the sample acted as project leaders in international projects at the moment of the study. All the students had five years of experience as software practitioners at the time the survey was applied. On the other hand, professors in the sample had five to eight years of experience in software development. Among them 30% were analysts, 28% programmers, 12% of the subjects had experience in testing and maintenance, 11% were designers, and the rest had been project leaders.

After the MBTI was applied, participants were queried about their preferences for the various roles: analyst, designer, programmer, tester, and maintainer. Specifically, participants were asked to state if they preferred, did not prefer, or were neutral about these roles. In particular, participants were instructed to focus only on their overall preferences and to disregard any particular software development tasks that they might have been performing at the moment the research was conducted.

## IV. RESULTS AND DISCUSSION

In the MBTI distribution of the 100 software developers (Table 2), specific poles dominate within each dimension. Specifically, the number of extroverts (63%) is almost double that of introverts (37%). Similarly, sensing individuals (72%) predominate over intuitive people (28%), thinking types (75%) are three times as common as feeling (25%), and judging people (61%) outnumber perceiving individuals (39%).

With reference to the individual poles, it is immediately clear that 'Ts' and 'Ss' (75% and 72% respectively) are over-represented in the sample, whereas 'Fs' and 'Ns' (25% and 28% each) are under-represented.

In terms of personality types, the ESTJ configuration is the most popular type (25%), while ESTP (15%) and ISTJ (10%) are also relatively common. In combination, these three personality types represent half of the sample. In contrast, the least represented combinations were INFJ and INFP, with each type indicating 1% of the subjects while ISFP, ENTP, and ESFJ only accounted for 2% of the total sample, as shown in Table II.

For statistical analysis, a Chi-Square (non-parametric) test was applied. The distribution of the observed results (Table II) was not found to be highly significant ($\alpha<0.001$), as compared to the expected distribution (Table I). Role preference data is presented in Table III. Each row depicts the personality type, the number of individuals comprising each category, and the number of individuals preferring each role. Among the roles: analysts, designers, and programmers are the most popular with analyst being the most preferred of all. In contrast, testers and maintainers were the least popular. The definitions of each role were shown to the subjects in a presentation before they were asked about their preferences.

TABLE II - CUBAN SOFTWARE ENGINEERS REPRESENTATION AMONG THE 16 MBTI TYPES (N= 100)

| ISTJ | ISFJ | INFJ | INTJ | E | I |
|---|---|---|---|---|---|
| 10% | 7% | 1% | 6% | 63% | 37% |
| ISTP | ISFP | INFP | INTP | S | N |
| 5% | 2% | 1% | 5% | 72% | 28% |
| ESTP | ESFP | ENFP | ENTP | T | F |
| 15% | 6% | 3% | 2% | 75% | 25% |
| ESTJ | ESFJ | ENFJ | ENTJ | J | P |
| 25% | 2% | 3% | 7% | 61% | 39% |

The temperaments that are certain about their preferences are SP/NF, set with the attitude pairs IP/EP, while the judging and orientation arises with FP/TP such as energy and perception point IN. On the other hand, SJ, TJ, and IS are less certain about the role that they want to assume in software development. Since extroverts are almost double the number of introverts in the sample, a comparison between those two poles is inappropriate.

*A. Analyst*

In order to check the statistical difference between the surveyed sample and the MBTI Type distribution among systems analysts, Table I was taken as the expected distribution of personality types for analyst, and those surveyed as observed results. A Statistical Binomial Test was performed for each 16 personality type presented in Table III, specifically those denoting preference for the role "Analyst." From these tests it can be inferred that: 21 (84%) of the individuals were from 25 ESTJ subjects, 14 (93%) individuals from 15 ESTP subjects, eight (80%) out of 10 ISTJ subjects, six (86%) out of seven ISFJ subjects, four (67%) out of six INTJs, six (100%) out of six ESFPs, and four (80%) out of five INTPs preferred the role "Analyst."

As expected, values of neutral and non-preference are not available, therefore generalized inferences could not be made. Although descriptive analysis framed on the sample shows: 43% ENTJs, 16% ESTJs, 10% ISTJs, and 7% ESTPs do not prefer the role "Analyst." In contrast, 14% ENTJs, 14% ISFJs, and 10% ISTJs express themselves as neutral about this role.

*B. Designer*

As there was no literature exposing what can be taken as expected values for this role, the authors proceed to describe preferences shown in Table III, aiming to stand with what in the near future can be taken as expected values. Several personality types within the surveyed sample prefer designing, including 100% INTPs, 83% INTJs, 80% ISTJs and ISTPs, 72% ESTJs, 71% ENTJs, 60% ESTPs, and 50% ESFPs. On the other hand, 33% ESFPs, 27% ESTPs, 29% ISFJs, 24% ESTJs, 20% ISTJs, 17% INTJs, and 14% ENTJs dislike designing. Lastly, 43% ISFJs, 17% ESFPs, 14% ENTJs, 10% ISTJs, 13% ESTPs, and 4% ESTJs state their neutrality about the role "Designer."

*C. Programmer*

The expected distribution of personality types for programmer has been presented in Table I. A Statistical Binomial Test was performed for each of the 16 personality types presented in Table III, specifically those denoting preference for the role Programmer. From these tests we inferred, with high significance, that: 19 (76%) individuals from 25 ESTJ subjects, 10 (67%) from 15 ESTP subjects, five (71%) out of seven ISFJ or ENTJ subjects, five (83%) individuals out of six ISFJ subjects, four (67%) out of six ESFPs, and five out of five ISTPs preferred programming.

Expected values of neutral and non-preference regarding programmers are not available; thus general inferences cannot be made in this case. Descriptive analysis framed on the sample indicates that: 60% ISTJs, 33.3% both ESTPs and ESFPs, 16.7% INTJs, 16% ESTJs, and 14.3% both ISFJs and ENTJs do not like programming. In contrast, 20% INTPs, 14% ISFJs, and ENTJs, 10% ISTJs, and 8% ESTJs declare their neutrality about programming.

*D. Tester*

Regarding tests activities the results are as follows: 30% ISTJs preferred to perform testing. In contrast, 83% INTJs, 80% ISTPs, 71% ENTJs, 67% ESFPs and ESTPs, and 57% ISFJs did not like testing. Personality types that reported neutral feelings for this role among the surveyed individuals included 30% ISTJs, 29% ENTJs, 20% ISTPs and INTPs, 17% ESFPs, 14% ISFJs, 13% ESTPs, and 4% ESTJs.

*E. Maintainer*

Finally, the following personality types preferred maintenance: 40% ISTPs, 29% ISFJs and ENTJs, 28% ESTJs, 20% ISTJs, 17% ESFPs, and 13% ESTPs. On the other hand, 83% ESFPs, 67% INTJs, 57% ISFJs, 53% ESTPs, 48% ESTJs, 43% ENTJs, 30% ISTJs, and 20% INTPs and ISTJs did not like maintenance. Lastly, 60% INTPs, 50% ISTJs, 40% ISTPs, 33% ESTPs, 33% INTJs, 29% ENTJs, 24% ESTJs, and 14% ISFJs expressed their neutrality about this role.

Once preferences have been related to each personality type, the following conclusions can be made about individual indicators, as presented in Table IV.

High preference percentages of the extroverts and introverts related to the analyst role can be noticed from Fig. 1. 84% out of 63 extroverts prefer to be analysts, 70% of introverts want to be designers, whereas 65% of introverts prefer doing programming. This is well in line with the suggestions by Capretz [13] and Capretz and Ahmed [28], who recommend that extroverts are more suitable to be appointed as system analysts. Their recommendations of introverts for the jobs of software designers and/or programmers are also supported by the outcome of this study.

In contrast to intuitive respondents, sensing individuals report higher preference for the design and/or maintenance activities respectively (63% over 39% regarding design and 24% over 14% regarding maintenance) as presented in Fig. 2.

TABLE III - SUMMARY OF MBTI ROLE PREFERENCES

| Types | Portion | % of Total | Analyst | | | Designer | | | Programmer | | | Tester | | | Maintainer | | |
|---|---|---|---|---|---|---|---|---|---|---|---|---|---|---|---|---|---|
| | | | 1 | 0 | -1 | 1 | 0 | -1 | 1 | 0 | -1 | 1 | 0 | -1 | 1 | 0 | -1 |
| ISTJ | 10 | 10 | 8 | 1 | 1 | 8 | 1 | 1 | 3 | 1 | 6 | 3 | 3 | 4 | 2 | 5 | 3 |
| ISFJ | 7 | 7 | 6 | 1 | 0 | 2 | 3 | 2 | 5 | 1 | 1 | 2 | 1 | 4 | 2 | 1 | 4 |
| INFJ | 1 | 1 | 1 | 0 | 0 | 1 | 0 | 0 | 1 | 0 | 0 | 0 | 0 | 1 | 0 | 0 | 1 |
| INTJ | 6 | 6 | 4 | 0 | 2 | 5 | 0 | 1 | 5 | 0 | 1 | 1 | 0 | 5 | 0 | 2 | 4 |
| ISTP | 5 | 5 | 2 | 1 | 2 | 4 | 0 | 1 | 5 | 0 | 0 | 0 | 1 | 4 | 2 | 2 | 1 |
| ISFP | 2 | 2 | 2 | 0 | 0 | 0 | 0 | 2 | 2 | 0 | 0 | 0 | 1 | 1 | 1 | 1 | 0 |
| INFP | 1 | 1 | 0 | 0 | 1 | 1 | 0 | 0 | 1 | 0 | 0 | 0 | 1 | 0 | 0 | 1 | 0 |
| INTP | 5 | 5 | 4 | 0 | 1 | 5 | 0 | 0 | 2 | 1 | 2 | 1 | 1 | 3 | 1 | 3 | 1 |
| ESTP | 15 | 15 | 14 | 0 | 1 | 9 | 2 | 4 | 10 | 0 | 5 | 3 | 2 | 10 | 2 | 5 | 8 |
| ESFP | 6 | 6 | 6 | 0 | 0 | 3 | 1 | 2 | 4 | 0 | 2 | 1 | 1 | 4 | 1 | 0 | 5 |
| ENFP | 3 | 3 | 2 | 0 | 1 | 1 | 1 | 1 | 3 | 0 | 0 | 0 | 0 | 3 | 0 | 0 | 3 |
| ENTP | 2 | 2 | 2 | 0 | 0 | 2 | 0 | 0 | 0 | 0 | 2 | 2 | 0 | 0 | 0 | 0 | 2 |
| ESTJ | 25 | 25 | 21 | 0 | 4 | 18 | 1 | 6 | 19 | 2 | 4 | 5 | 1 | 19 | 7 | 6 | 12 |
| ESFJ | 2 | 2 | 2 | 0 | 0 | 1 | 0 | 1 | 1 | 0 | 1 | 1 | 0 | 1 | 0 | 0 | 2 |
| ENFJ | 3 | 3 | 3 | 0 | 0 | 2 | 1 | 0 | 0 | 1 | 2 | 1 | 0 | 2 | 1 | 0 | 2 |
| ENTJ | 7 | 7 | 3 | 1 | 3 | 5 | 1 | 1 | 5 | 1 | 1 | 0 | 2 | 5 | 2 | 2 | 3 |

Note: 1 means "prefer," 0 means "neutral," and -1 means "do not prefer."

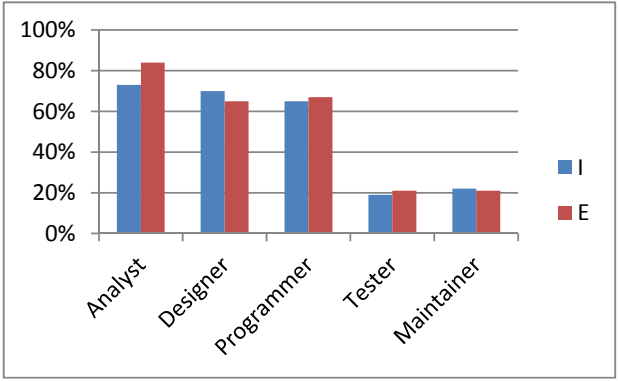

Fig. 1. E-I Preference Indicator

While mapping skills to MBTI personality types, Capretz and Ahmed [13] also found intuitive individuals more likely to thrive in design and programming, which is also supported by this study's results. However, they believe sensing people are more suitable for testing and maintenance jobs. In contrast, only 21% of sensing individuals showed their preference for testing and 24% of them wanted to do software maintenance. Furthermore, a high percentage of both sensing and intuitive individuals opted to work as analysts. This came as a surprise, if we compare it with the observation of other researchers [13], [22].

Capretz and Ahmed [13] assert that Ts are not a preferable choice for software analyst's job; they are more likely to prefer programming. In this study, however, there is a remarkable preference for designers (75%) in the thinking individuals as compared to only 44% by the feeling individuals. It is, however, notable that feeling individuals express (11%) more preferences for analyst. The interest upon the rest of the roles remains almost equal for both dimensions, as shown in Fig. 3.

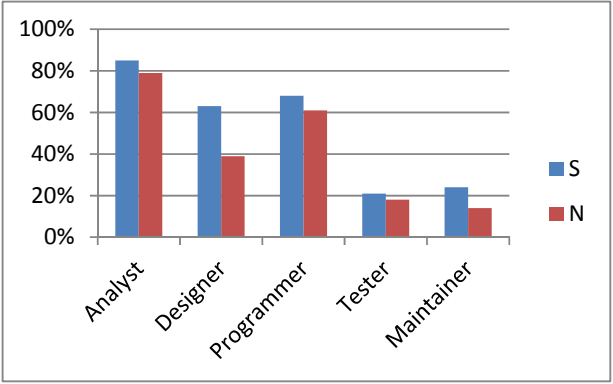

Fig. 2. S-N Preference Indicator

In the judging and perceiving dimensions, there was not much difference with regards to role preference percentages. However, the high preference for the role of analyst by both judging (79%) and perceiving (82%) is remarkable. This is followed by their preference for the roles of designers and programmers, as presented in Fig. 4. This is surprisingly in conflict with the theoretical outcome, according to which sensing and judging individuals are more suitable for software testing [13]. This result is also confirmed by the work of Tanjila et al. [29].

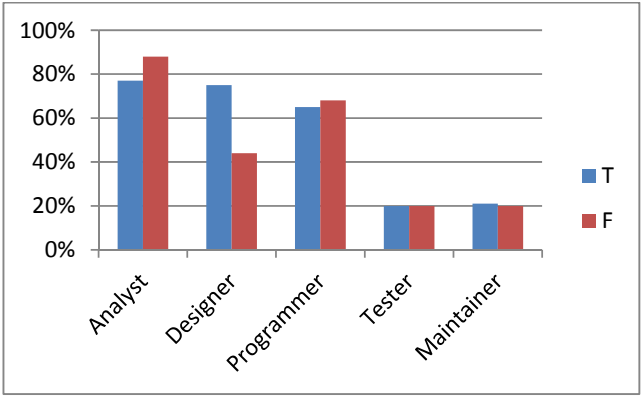

Fig. 3. T-F Preference Indicator

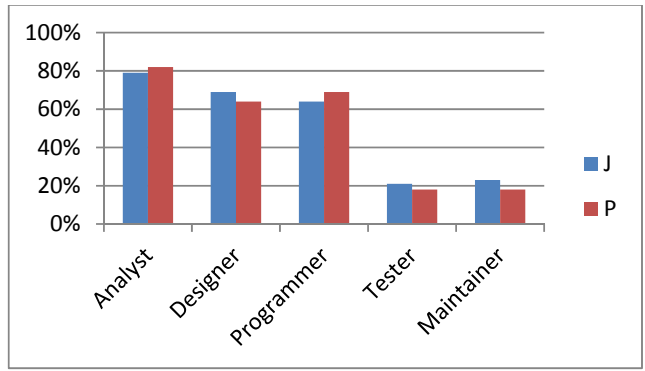

Fig. 4. J-P Preference Indicator

The predominant temperament, as can be seen in Table IV, within the sample is SJ, with 45% of individuals preferring this type. In contrast, the least popular temperament is NF, with only 8% of people having these traits. The temperament results are similar to those in previous studies [24] and [30]. However, SP was the second most represented temperament type, rather than NT, which was the case in those studies.

TABLE IV - INDIVIDUAL INDICATOR AND TEMPERAMENT DISTRIBUTION

| Types | Quantity | Analyst | | Designer | | Programmer | | Tester | | Maintainer | |
|---|---|---|---|---|---|---|---|---|---|---|---|
| I | 37 | 73% | 27 | 70% | 26 | 65% | 24 | 19% | 7 | 22% | 8 |
| E | 63 | 84% | 53 | 65% | 41 | 67% | 42 | 21% | 13 | 21% | 13 |
| S | 72 | 85% | 61 | 63% | 45 | 68% | 49 | 21% | 15 | 24% | 17 |
| N | 28 | 79% | 22 | 39% | 11 | 61% | 17 | 18% | 5 | 14% | 4 |
| T | 75 | 77% | 58 | 75% | 56 | 65% | 49 | 20% | 15 | 21% | 16 |
| F | 25 | 88% | 22 | 44% | 11 | 68% | 17 | 20% | 5 | 20% | 5 |
| J | 61 | 79% | 48 | 69% | 42 | 64% | 39 | 21% | 13 | 23% | 14 |
| P | 39 | 82% | 32 | 64% | 25 | 69% | 27 | 18% | 7 | 18% | 7 |
| **Temperament** | | | | | | | | | | | |
| SP | 27 | 85% | 23 | 59% | 16 | 74% | 20 | 11% | 3 | 22% | 6 |
| SJ | 45 | 84% | 37 | 66% | 29 | 64% | 28 | 25% | 11 | 25% | 11 |
| NT | 20 | 65% | 13 | 85% | 17 | 60% | 12 | 20% | 4 | 15% | 3 |
| NF | 8 | 75% | 6 | 63% | 5 | 63% | 5 | 13% | 1 | 13% | 1 |

The most highly-valued preference for the SP temperament was the analyst, which was followed by the programmer role. Furthermore, the NTs have a clear preference for the designer role, while the SJ and NF types preferred the role "Analyst." Lastly, the NFs preferred the roles "Designer" and "Programmer."

## V. CONCLUSION

An empirical study was conducted to map some opposing psychological traits, such as extroversion-introversion, sensing-intuition, thinking-feeling, and judging-perceiving to the main tasks of a software life cycle. This work addresses the fundamental issue of the impact of human factors in software development. This research provides an empirical assessment of the relationships between software engineers' MBTI types and role preferences. Although the MBTI does not predict success in a particular career, it identifies individual preferences for specific occupations.

The results indicated by the personality types INFJ, ISFP, INFP, ESFP, ENFP, ENTP, ESFJ, and ENFJ are not relevant because of the sample size. In each of these cases, our data values are in the range of one and three individuals, thus it lacks statistical significance.

However, distinctive patterns are evident in the relationship between personality types and preferences of software engineering roles. This study revealed particular role preferences for each personality type. It has been inferred, with high significance, that 84% ESTJs, 93% ESTPs, 80% ISTJs, 86% ISFJs, 67% INTJs, 100% ESFPs, and 80% INTPs indicate analysis in their preferences. While 76% ESTJs, 67 ESTPs, 71% ISFJs/ENTJs, 83% ISFJs, 67% ESFPs, and 100% ISTPs prefer programming. On the other hand, preferences for designing, testing, and maintenance are described according to the sample. It can be concluded that assigning a person with specific psychological characteristics to tasks of the software life cycle best suited for his/her traits increases the chances of a successful outcome for the project.

Overall, the roles of analyst, designer, and programmer were the most popular roles among all personality types, whereas tester and maintainer were identified as the least desirable. The margins of neutral positions can be decreased through further inquiries about software engineer preferences.

Finally, it is important to point out that the study of the relations between personality types and software roles provides more light as to how software development is affected based on individual type indicators and how important it is to match the right people to roles in software engineering.